\documentclass[12pt]{article}
\usepackage{amssymb,epsfig}
\newcommand{\es}{$\mathrm{E_6}$}

\begin{document}

\hspace*{\fill}UWThPh-2001-45

\hspace*{\fill}WUE-ITP-2001-034

\hspace*{\fill}hep-ph/0111285

\vspace{1cm}

\begin{center}
\textbf{\Large Production of singlino dominated neutralinos in
extended supersymmetric models}

\vspace{10mm}
F.~Franke\footnote{e-mail: fabian@physik.uni-wuerzburg.de} 

Institut f\"ur Theoretische Physik und Astrophysik, 
Universit\"at W\"urzburg,\\
D-97074 W\"urzburg, Germany

\vspace{5mm}
S.~Hesselbach\footnote{e-mail: hesselb@thp.univie.ac.at} 

Institut f\"ur Theoretische Physik, Universit\"at Wien,
A-1090 Wien, Austria

\end{center}

\vspace{5mm}
\begin{abstract}
Neutralinos with a large singlino component may appear
in extended supersymmetric models with additional singlet Higgs fields.
Since singlinos do not couple to (s)fermions and gauge bosons, the 
cross sections for the production of singlino dominated neutralinos are
generally small. Within the framework of the Next-to-Minimal Supersymmetric
Standard Model (NMSSM) and an \es\ inspired model we study neutralino
production 
$e^+e^- \rightarrow \tilde{\chi}^0_1 \tilde{\chi}^0_2$
($\tilde{\chi}^0_{1,2} \tilde{\chi}^0_3$) with a singlino dominated
$\tilde{\chi}^0_2$ ($\tilde{\chi}^0_3$). 
It is shown that neutralinos with a singlino contribution up to 99\%
can be produced with a cross section larger than 1 fb and may therefore
be detected at a high luminosity $e^+e^-$ linear collider even if they are
not the LSP. 
\end{abstract}

\section{Introduction}

The production of neutralinos, the supersymmetric partners of
the neutral gauge and Higgs bosons, is expected to be among the most 
promising processes to discover a supersymmetric particle.
As soon 
as a neutralino is detected a detailed study of 
its mass and mixing properties is necessary 
in order to deduce the underlying supersymmetric model.
It has to be determined whether a neutralino can be assigned to the
Minimal Supersymmetric Standard Model (MSSM) or belongs to 
an extension by additional singlet fields.
A characteristic feature of extended supersymmetric models
with singlet Higgs fields is the existence of neutralinos with large singlino
components. Since, however, their couplings to (s)fermions and gauge bosons
are strongly suppressed, it is often expected that only a singlino dominated
neutralino which is the lightest supersymmetric particle (LSP)
can be identified in extended models with $R$-parity conservation where it
is in the final state in all supersymmetric decay chains \cite{hugonie}.  
Especially within the framework of the Next-to-Minimal Supersymmetric 
Standard Model (NMSSM) it is argued that a singlino-like neutralino which
is not the LSP could not be detected since it is not directly  
produced with a sufficient cross section and also 
omitted in the decay chains 
of the heavier supersymmetric particles.

Consequently, nonminimal neutralino production so far has only been
discussed in supersymmetric scenarios where the lightest
neutralino has a dominant singlet component and the other neutralinos
have mainly MSSM properties \cite{singlinoprod, singlinoprodpol}. 
The aim of this study is a detailed analysis of the
production cross section
$e^+e^- \rightarrow \tilde{\chi}^0_1 \tilde{\chi}^0_2$
($\tilde{\chi}^0_1 \tilde{\chi}^0_3$, 
$\tilde{\chi}^0_2 \tilde{\chi}^0_3$), 
with a singlino dominated second (third) neutralino. 

In Sec.~2  our scenarios in two 
representative extended supersymmetric models,
the unconstrained 
NMSSM and an \es\ model with an additional neutral gauge boson
are described. A detailed analysis of the cross sections including
the option of polarized electron and positron beams 
as a function of the singlino component of the produced neutralino
follows in
Sec.~3.
In our conclusion we point out that, depending on
the supersymmetric parameters and the selectron masses, cross
sections up to 1 fb can be reached even for neutralinos with a 
singlino content of 99\%. Therefore a high luminosity 
linear collider with
polarized beams promises to be an excellent tool for the detection of
singlino dominated non-LSP neutralinos and a discrimination between the 
minimal and extended supersymmetric models.

\section{Scenarios}

The production of a singlino dominated
neutralino with a mass of 150 GeV that is not the LSP 
at a linear $e^+e^-$collider is discussed within the framework of 
two extensions of the Minimal Supersymmetric 
Standard Model (MSSM) by a singlet Higgs
field $S$.

\subsection{NMSSM}

In the unconstrained Next-to-Minimal Supersymmetric Standard Model
(NMSSM) \cite{nmssm1}
the masses and couplings of the five neutralinos 
depend on the the gaugino 
mass parameters $M_2$ and $M_1$, the ratio of the vacuum 
expectation values of the doublet Higgs fields $H_1$ and $H_2$,
$\tan\beta = v_2/v_1$,
the vacuum expectation value $x$ of the singlet Higgs field $S$ and the 
trilinear couplings $\lambda$ and $\kappa$ in the superpotential
$W \supset \lambda H_1 H_2 S - 1/3 \kappa S^3$
\cite{nmssm2}.
In our calculations we always assume the GUT relation $M_1/M_2= 5/3
\tan^2 \theta_W$.
For large $x \gg |M_2|$ a singlino dominated neutralino with mass
eigenvalue $\approx 2 \kappa x$ in zeroth approximation decouples in 
the neutralino mixing matrix
while the other neutralinos have MSSM character with
$\mu = \lambda x$. The singlino content of the $i$-th neutralino 
$\tilde{\chi}_i^0$
is described by the squared matrix element $|N_{i5}|^2$
of the unitary $5 \times 5$ matrix $N$
which diagonalizes the neutralino mass matrix 
in the basis
$(\tilde{\gamma},\tilde{Z},\tilde{H_1},\tilde{H_2},
\tilde{S})$ of the photino, zino, the two doublet higgsinos and the
singlino. 

In the following section we analyze the cross sections for the production of
a singlino dominated neutralino together
with an MSSM-like $\tilde{\chi}_1^0$ or $\tilde{\chi}_2^0$ in  
representative scenarios where the LSP
is mainly a gaugino or a higgsino with mass 100~GeV
(Table 1). 
The singlet vacuum expectation value $x$ is varied in order to
study different singlino contents, but 
the mass of the 
singlino dominated neutralino 
which is the second lightest neutralino in the gaugino scenario
NMSSM-G
($M_2=211$ GeV, $\lambda x=400$ GeV) and the third lightest in the higgsino
scenario NMSSM-H ($M_2=-400$ GeV, $\lambda x=107$ GeV) 
is fixed at 150~GeV by the parameter $\kappa$. 
Note that in the higgsino scenario both 
light neutralinos have higgsino character
with different signs of the mass eigenvalues.

Fig.~\ref{sing} shows the mixing character of the singlino dominated
neutralinos.
The singlino component crucially depends on the singlet
vacuum expectation value $x$ \cite{displaced}. 
In order to obtain a singlino content of 90\%, $x$ must be
larger than 1200 GeV (NMSSM-G) or 1000~GeV (NMSSM-H).
Generally the singlino component increases with increasing value of
$x$.

A singlino content $|N_{25}|^2$ of 99\% for
$\tilde{\chi}_2^0$ with
$m_{\tilde{\chi}_2^0}=150$ GeV is reached at $x=3750$~GeV, $\kappa=0.02$
in the gaugino scenario.
In the higgsino scenario $\tilde{\chi}_3^0$ is a 99\% singlino for
$x=2750$ GeV and $\kappa=0.0275$ (Table 1). In section 3 these values
will be used to compute production cross sections.

The couplings of the singlino dominated neutralino and therefore 
its production cross sections are determined by the remaining MSSM
components. It is the gaugino component 
that dominates the MSSM content in the scenario NMSSM-G
with a gaugino-like LSP while in the scenario NMSSM-H with a higgsino-like
LSP the MSSM-higgsino component prevails. 
These mixing characteristics facilitate the identification of the production
of the singlino-like neutralino since it enhances the corresponding
cross sections for the associated production of the light neutralinos
(Sec.~3).

\begin{table}[t]
\begin{center}
\renewcommand{\arraystretch}{1.3}
\begin{tabular}{|c|c|c|c|c|c|c|}
\hline
Scenario & NMSSM-G & NMSSM-H & \es-G$+$ & \es-G$-$ & \es-H$+$ & \es-H$-$ \\
\hline \hline
$M_2/$GeV & 211 & $-400$ & 211 & 211 & $-400$ & $-400$ \\
$\lambda x$/GeV & 400 & 107 & 400 & 400 & 107 & 107 \\
$\tan\beta$ & 3 & 3 & 3 & 3 & 3 & 3\\ 
$\tilde{\chi}_1^0$ character & gaugino & higgsino & gaugino & gaugino &
higgsino & higgsino \\ 
$m_{\tilde{\chi}_1^0}/$GeV & (+) 100 & (+) 100 & (+) 100 & (+) 100 & 
(+) 100 & (+) 100 \\
$m_{\tilde{\chi}_2^0}/$GeV & (+) 150 & ($-$) 122 & ($-$) 150 & (+) 150 &
($-$) 122 & ($-$) 122\\
$m_{\tilde{\chi}_3^0}/$GeV & (+) 190 & ($-$) 150 & (+) 190 & (+) 190 &
(+) 150 & ($-$) 150\\
\hline\hline
$x$/GeV & 3750 & 2750 & 4100 & 4700 & 3700 & 4200 \\  
$m_{Z'}/GeV$ & -- & -- & 1726 & 1979 & 1558 & 1768 \\
$\kappa$ & 0.02 & 0.0272 & -- & -- & -- & --\\
$M'$ &  -- & -- & 19652.4 & $-25979.6$ & 16015.4 & $-20704.8$ \\
$|N_{25}|^2$ & 0.99 & & 0.99 & 0.99 & & \\ 
$|N_{35}|^2$ &  & 0.99 & & & 0.99 & 0.99 \\
\hline\hline
\end{tabular}
\end{center}
\caption{\label{scentab}Parameters of the supersymmetric models in
  representative scenarios.   
  The mass of the singlino dominated neutralino is fixed by the parameters
  $\kappa$ (NMSSM) and $M'$ (\es\ model). The signs in parentheses
  show the relative sign of the neutralino mass eigenvalues in
  comparison to the mass eigenvalues of the lightest neutralino.}
\end{table}

\subsection{ $\mathbf{E_6}$ model}
We consider an \es\ model with one extra neutral gauge boson $Z'$
and one additional singlet superfield 
\cite{e6model}.
To respect the experimental mass bounds for new gauge bosons
the singlet vacuum expectation value $x$ must be
larger than about 1.5~TeV~\cite{abe}.
The extended neutralino sector in this model contains six
neutralinos being mixtures of photino, zino,
doublet higgsinos, $Z'$ gaugino ($\tilde{Z}'$), and singlino
\cite{e6neutralino, hesselb}. The $6 \times 6$ neutralino mass matrix
depends on six parameters: the $\rm SU(2)_L$, U(1)$_Y$ and
$\rm U(1)'$ gaugino
mass parameters $M_2$, $M_1$ and $M'$, $\tan\beta$, $x$
and the trilinear coupling $\lambda$ in the superpotential
$W \supset \lambda H_1 H_2 S$.
Analogously to the NMSSM, the matrix element 
$|N_{i6}|^2$ of the unitary $6 \times 6$ matrix that
diagonalizes the neutralino mass matrix describes the
singlino content of the $i$-th neutralino in the basis 
$(\tilde{\gamma},\tilde{Z},\tilde{H_1},\tilde{H_2},\tilde{Z}',\tilde{S})$.

In the \es\ model 
the $4\times 4$ submatrix of the MSSM-like neutralinos and the
$2\times 2$ submatrix of the exotic ones approximately decouple
because of the large values of $x$.
For very large values of $|M'| \gg x$  
a neutralino with a mass of 150 GeV
can be a nearly pure singlino \cite{decarlosespinosa}
with mass $\approx 0.18 \,x^2/|M'|$ in zeroth approximation.

As in the NMSSM we choose scenarios where the lightest neutralino
with a mass of 100~GeV is mainly a gaugino or a higgsino (Table 1).
The singlino-like neutralino 
($\tilde{\chi}^0_2$ in the gaugino scenarios and
$\tilde{\chi}^0_3$ in the higgsino scenarios)
can be fixed at 150~GeV by choosing an appropriate 
value for $M'$ even if the singlet vacuum expectation value
is varied.
For positive $M'$ the singlino dominated neutralino has opposite
sign of the mass eigenvalue relative to the LSP than in the NMSSM,
while for $M'<0$ the mass eigenvalue of this neutralino flips sign and
thus becomes the same as in the NMSSM.
Therefore the properties of the singlino-like neutralino
in the \es\ scenarios \es-G$-$ and \es-H$-$ with negative $M'$ are more
similar to those in the NMSSM.

The MSSM components of the singlino dominated neutralino
in the \es\ scenarios
are, however, significantly smaller than the
$\tilde{Z}'$ components and also than the MSSM components in the NMSSM.
This mixing characteristic
leads to smaller production cross sections compared to the NMSSM
as discussed in the next section. Contrary to the NMSSM, in the
scenario \es-G+ with a gaugino-like LSP the MSSM doublet higgsino component
even outweighs the MSSM gaugino component which results in a further
suppression of the production cross section in this scenario.

\begin{figure}[p]
\begin{picture}(16,17.7)
%\put(0,0){\framebox(16,17.7){}}

\put(0,0){\epsfig{file=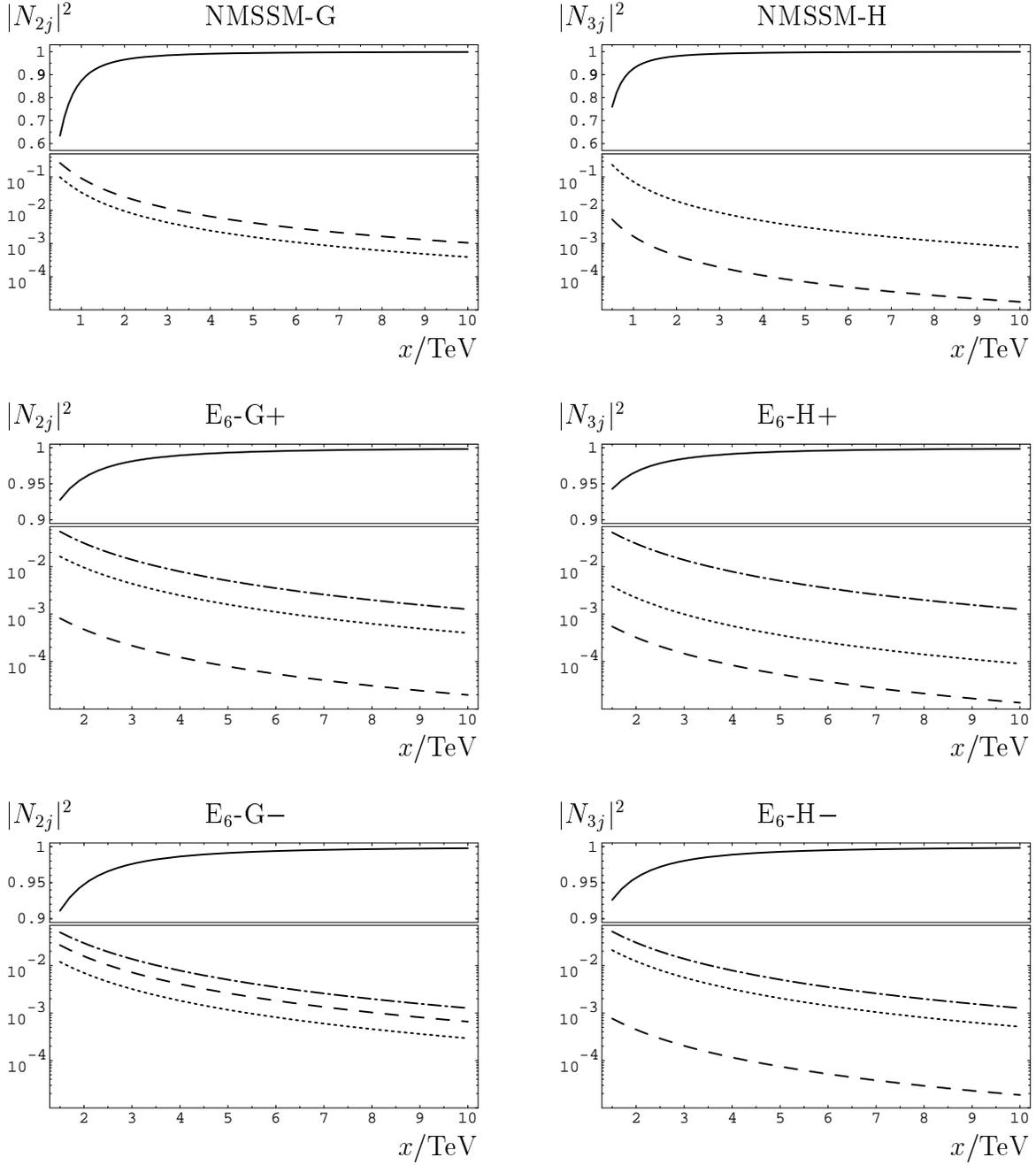}}

\end{picture}
\caption{\label{sing}Mixing of the singlino dominated
neutralino in the scenarios of Table 1:
singlino component (solid line),
MSSM gaugino components ($|N_{i1}|^2 + |N_{i2}|^2$, dashed),
MSSM doublet higgsino components ($|N_{i3}|^2 + |N_{i4}|^2$, dotted)
and \es\ $\tilde{Z}'$ component (dashed-dotted).
The mass of the singlino dominated neutralino is fixed at 150~GeV by
the parameters $\kappa$ (NMSSM) and $M'$ (\es\ model).}
\end{figure}

As in the NMSSM the singlino content increases also in the \es\ model with
increasing vacuum expectation value $x$ (Fig.~\ref{sing}). 
In order to obtain a singlino content exceeding
99\% the singlet vacuum expectation value 
$x$ must be larger than about 3.5 -- 4 TeV in our scenarios. 

\section{Cross sections}
We study the associated production of a singlino dominated neutralino
$e^+e^- \rightarrow \tilde{\chi}^0_1 \tilde{\chi}^0_2$ in the
scenarios with a gaugino-like LSP and 
$e^+e^- \rightarrow \tilde{\chi}^0_{1,2} \tilde{\chi}^0_3$ in the
scenarios where the lightest and second lightest neutralinos
are mainly higgsinos.
Neutralino production in $e^+e^-$ annihilation proceeds via
the exchange of a $Z$ boson in the $s$ channel and of left and
right selectrons in the $t$ and $u$ channel. In the
\es\ model the contribution of the additional $Z'$ boson has to
be considered. Due to the large $Z'$ mass in the scenarios of
Table 1, however, $Z'$ resonance effects appear far beyond expected
energies of a linear collider at first stages.
Since the singlino component does not couple to gauge bosons
and fermions or sfermions due to the hypercharge 0 of the singlet field,
the cross section for the production of neutralinos with a 
significant singlino 
component is strongly suppressed.

Analytical formulae are given in \cite{franke} for the NMMSM and
\cite{hesselb} for the \es\ model.
For comparison, the masses of the left and right selectrons
$m_{\tilde{e}_L}=300$ GeV, $m_{\tilde{e}_R}=200$ GeV
are fixed in all scenarios.

The use of polarized beams may be an excellent tool to 
enhance cross sections and to identify the supersymmetric signal.
Polarizations up to 85\% for electrons and 60\% for positrons are
expected at a future linear collider \cite{pol}.
We consider the polarization configurations 
$P_-=+0.85$, $P_+=-0.6$ (right polarized electrons 
and left polarized positrons) and
$P_-=-0.85$, $P_+=+0.6$ (left polarized electrons 
and right polarized positrons) 
in our studies.

In the following we assume a cross section of 1 fb to be
sufficient for the identification of a neutralino production
process. Of course the discovery limit depends on
the neutralino decay properties that are discussed in detail
in \cite{hugonie, displaced, franke}.

\begin{figure}[p]
\begin{picture}(16,17.7)
%\put(0,0){\framebox(16,17.7){}}

\put(0,0){\epsfig{file=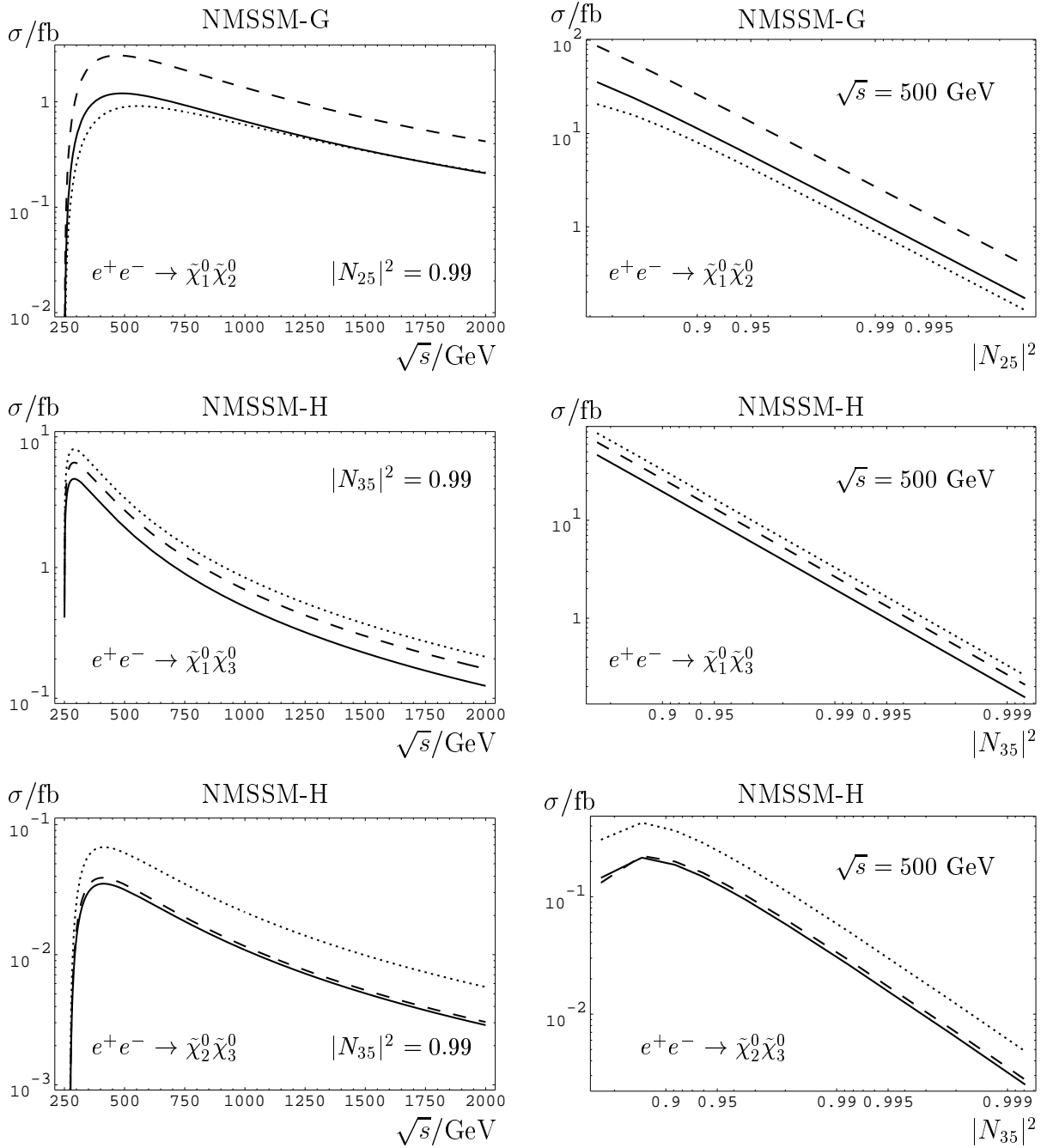}}

\end{picture}
\caption{\label{NMSSM}Cross sections for the production of a 
singlino dominated neutralino in the NMSSM scenarios given in Table 1
with unpolarized beams (solid) and beam polarizations
$P_-=+0.85$, $P_+=-0.6$ (dashed) and
$P_-=-0.85$, $P_+=+0.6$ (dotted).
In the plots on the right hand side the singlet vacuum expectation
value $x$ is varied and the mass of the singlino dominated neutralino is
fixed at 150~GeV by the parameter $\kappa$.}
\end{figure}

In Fig.~\ref{NMSSM}  
the cross sections for the production of a neutralino with
a singlino content of 99\% are shown for beam energies
between the threshold and 2 TeV in the NMSSM scenarios of Table 1. 
For unpolarized beams the cross section for
$\tilde{\chi}^0_1 \tilde{\chi}^0_2$ production reaches values up
to 1 fb in the gaugino scenario NMSSM-G with selectron exchange
being the dominant channel.

In the higgsino scenario NMSSM-H neutralino production proceeds
mainly by $Z$ boson exchange.
Then the cross sections for
$e^+e^-\rightarrow \tilde{\chi}^0_{1,2} \tilde{\chi}^0_3$
of the production of a singlino dominant $\tilde{\chi}^0_3$ 
depend on the MSSM-higgsino content of $\tilde{\chi}^0_3$ 
and the $Z$-higgsino coupling. 
While $\tilde{\chi}^0_1 \tilde{\chi}^0_3$ production reaches unpolarized cross
sections up to 5 fb, the cross section for
$\tilde{\chi}^0_2 \tilde{\chi}^0_3$ is nearly two orders of magnitude
smaller and significantly below discovery limit.
In the latter case the ratio of the doublet higgsino components
$N_{33}/N_{34}$ of $\tilde{\chi}^0_3$ is similar to
$N_{23}/N_{24}$ of $\tilde{\chi}^0_2$ which results in a small
$Z$-higgsino coupling because higgsino pair production is
generally suppressed.

The unpolarized cross section can be significantly enhanced
by a factor of 2 -- 3
in all scenarios by choosing a suitable beam polarization. 
In the gaugino scenario NMSSM-G 
where the neutralinos $\tilde{\chi}^0_1$ and $\tilde{\chi}^0_2$ couple
preferably to the right selectron,
the polarization configuration
$P_->0$, $P_+<0$ enlarges the cross sections, while in the
higgsino scenario the cross sections are largest for
$P_-<0$, $P_+>0$ \cite{singlinoprodpol, gudi}. 

Fig.~\ref{NMSSM} also shows the cross section for the associated
neutralino production at a linear collider with a center-of-mass
energy of 500 GeV as a function of the singlino content.
For the associated production of the 
second lightest neutralino which is a 90\% singlino 
and a gaugino-dominated LSP in scenario NMSSM-G one obtains 
cross sections of 15 fb with unpolarized
beams or 40 fb with beam configuration 
$P_-=+0.85$, $P_+=-0.6$.
In scenario NMSSM-H the third lightest neutralino with a singlino
content of 90\%
and the higgsino-dominated LSP are produced with 20~fb (unpolarized beams) and 
40~fb ($P_-=-0.85$, $P_+=+0.6$) in the
process $e^+e^-\rightarrow \tilde{\chi}^0_1 \tilde{\chi}^0_3$.
Assuming a discovery limit of 1 fb, a neutralino with a singlino 
content of 99.2\% can be detected at a linear collider 
with unpolarized beams in the 
gaugino scenario and even with a singlino content of 99.5\%
in the higgsino scenario although it is not the LSP.

\begin{figure}[p]
\begin{picture}(16,17.7)
%\put(0,0){\framebox(16,17.7){}}

\put(0,0){\epsfig{file=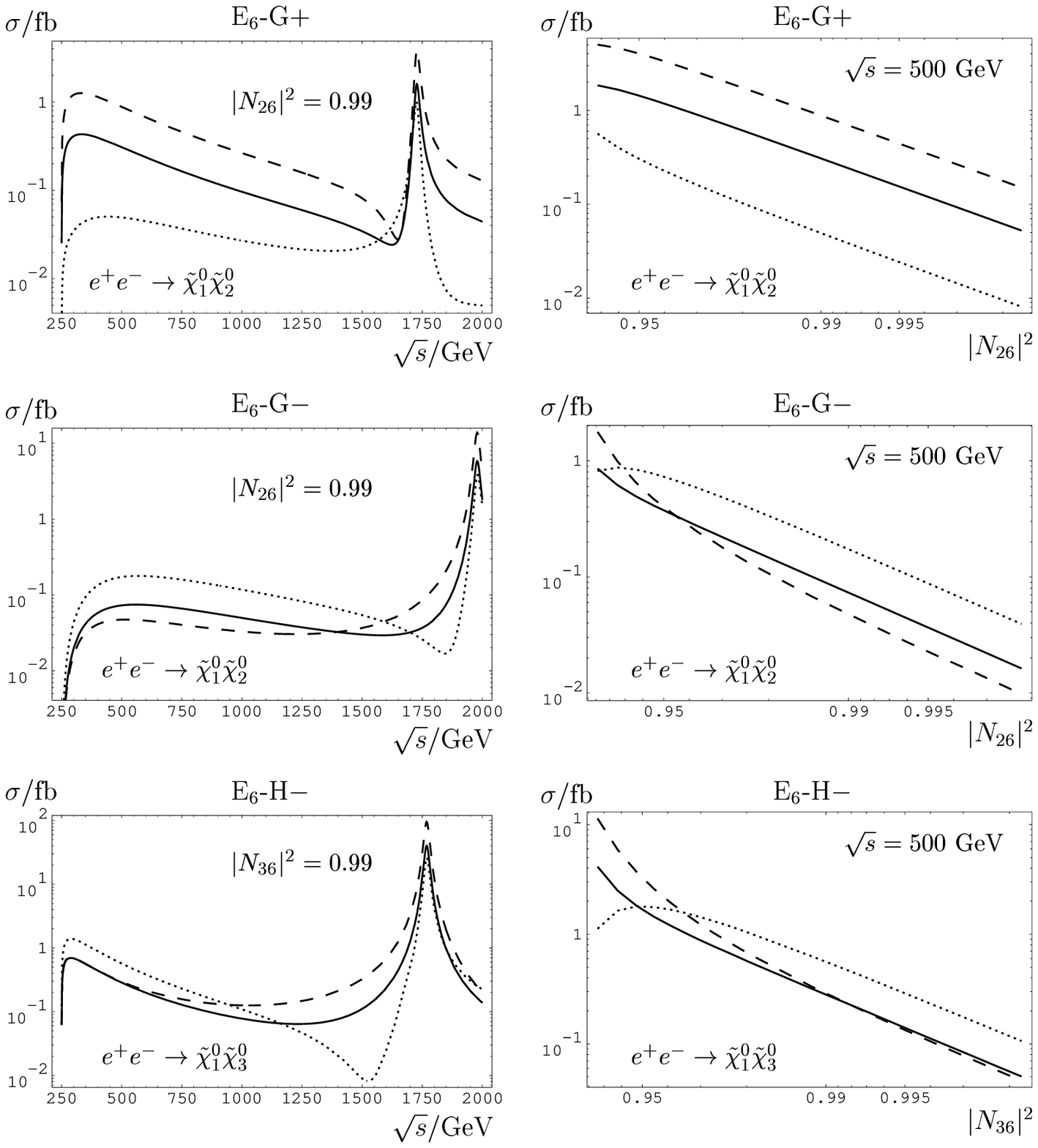}}

\end{picture}
\caption{\label{E6}Cross sections for the production of a 
singlino dominated neutralino in the \es\ model scenarios given in Table 1
with unpolarized beams (solid) and beam polarizations
$P_-=+0.85$, $P_+=-0.6$ (dashed) and
$P_-=-0.85$, $P_+=+0.6$ (dotted).
In the plots on the right hand side the singlet vacuum expectation
value $x$ is varied and the mass of the singlino dominated neutralino is
fixed at 150~GeV by the parameter $M'$.}
\end{figure}

In Fig.~\ref{E6} we show the corresponding cross sections 
in the \es\ model for the associated 
production of a singlino-dominated neutralino 
in the \es\ scenarios
of Table~\ref{scentab}.
In the \es\ model the singlino-dominated neutralino also has a
considerable $\tilde{Z}'$ component which even is the second largest
component.
It couples to the selectrons exchanged in 
$t$ and $u$ channels \cite{hesselb} but nevertheless
contributes significantly less
to the cross section than the doublet higgsino and gaugino components.
Therefore the cross sections for the associated production of a 
singlino-dominated neutralino are in the \es\ model 
for energies below the $Z'$ resonance 
generally smaller than in the NMSSM.

In scenario
\es-G+ both the singlino-dominated $\tilde{\chi}^0_2$ 
and the gaugino-like LSP couple preferably to the right
selectron. The unpolarized cross section for the associated production
of $\tilde{\chi}^0_2$ with a singlino content of 99\%
reaches a maximum value
of 0.4~fb for $\sqrt{s}<1500$~GeV which is enhanced to
1.2~fb 
for beam polarization $P_-=+0.85$, $P_+=-0.6$.
Cross sections of 1~fb at a linear collider with $\sqrt{s}=500$~GeV
are obtained for the associated production of a
neutralino with a singlino content as high as 97\% 
(unpolarized beams) or 99\% ($P_-=+0.85$, $P_+=-0.6$).
%in the \es\ scenario with a gaugino-like LSP and $M'>0$.

In scenario \es-G$-$, however, the singlino-dominated
$\tilde{\chi}^0_2$ couples
mainly to the heavier left selectron contrary to
$\tilde{\chi}^0_1$. Hence the cross sections are considerably smaller
with only 0.07~fb at $\sqrt{s}=500$~GeV for unpolarized beams.
Here a second lightest
neutralino with a singlet component smaller than 93\%
is produced 
with a cross section of at least 1 fb 
at a 500~GeV linear collider with beam polarization
$P_-=+0.85$, $P_+=-0.6$. Due to the increased contribution of 
the exchange of the left selectron it is the opposite polarization
configuration $P_-=-0.85$, $P_+=+0.6$
compared to scenario \es-G+
that enhances the cross section for larger singlino components. 

As to the \es\ higgsino scenarios only cross sections for the process $e^+e^-
\to \tilde{\chi}^0_1 \tilde{\chi}^0_3$ in scenario \es-H$-$
are given in Fig.~\ref{E6}. 
In scenario \es-H+ the cross sections for the associated production
of the singlino-dominated $\tilde{\chi}^0_3$ are strongly suppressed
since its doublet higgsino content is
more than one order of magnitude smaller
than in the NMSSM. 
Similar to the NMSSM also the $\tilde{\chi}^0_2$$\tilde{\chi}^0_3$
production in scenario 
\es-H$-$ is invisible.
However, 
the unpolarized cross section for $\tilde{\chi}^0_1
\tilde{\chi}^0_3$ production at a linear collider
reaches 0.7~fb outside the $Z'$ resonance 
and is enhanced to 1.4~fb with beam polarization $P_-=-0.85$,
$P_+=+0.6$.

Assuming a discovery limit of 1 fb, a neutralino with a singlino 
content up to 96.5\% (unpolarized beams) or even up to
98.2\% ($P_-=-0.85$, $P_+=+0.6$) can be detected at a linear collider 
with $\sqrt{s}=500$ GeV in the higgsino scenario \es-H$-$. 

Generally the beam polarization configurations 
which enhance the cross sections in gaugino and higgsino scenarios 
in the \es\ model outside the $Z'$ resonance are independent of the
energy or the singlino content.
In the vicinity of the $Z'$ resonance, however, resonance effects
radically change the polarization asymmetry. In our scenarios
these effects 
are out of reach at the first stages of a linear collider, but appear
at $\sqrt{s}=500$ GeV for a singlino content less than 96\% in the scenario 
\es-G$-$ and 97\% in 
\es-H$-$, where the $Z'$ boson becomes rather light with a mass
smaller than 1~TeV.

\section{Conclusion}

Experimental evidence of a neutralino with a significant singlino component
would be an explicit proof for 
an extended supersymmetric model with additional Higgs singlet fields.
A high luminosity $e^+e^-$ linear collider offers the opportunity
to study the associated production of a singlino-dominated
neutralino.
Complementary to existing
analyses we have shown that 
assuming a discovery limit of 1 fb
a neutralino with a singlino content larger than 90\% can be
detected at a linear collider
although it is not the LSP.
The use of 
suitably polarized beams
enhances the production cross sections and
facilitates the identification of the singlino-dominant
neutralino.

In representative scenarios with a gaugino-like
LSP one obtains unpolarized cross sections of 1 fb for the 
associated production 
$e^+e^- \rightarrow \tilde{\chi}^0_1 \tilde{\chi}^0_2$
of a 
second lightest neutralino with a singlino content of 99.2\%
in the NMSSM and 97\% in the \es\ model at a linear collider
with 500 GeV center-of-mass energy. Due to the additional
$\tilde{Z}'$ component of the neutralinos, cross sections are
generally smaller in the \es\ model below 
the $Z'$ resonance. In gaugino scenarios
mainly the exchange of left and right selectrons contribute
to the cross section. 
If the coupling to right (left) selectrons is dominant 
the beam polarization $P_->0$, $P_+<0$ ($P_-<0$, $P_+>0$)
increases the cross section by a factor of 2 $-$ 3. 

Also in scenarios 
where both light neutralinos have higgsino character
the cross sections for the associated production of
a singlino-dominated third neutralino
exceed the discovery limit. Here it is
the polarization configuration
$P_-<0$, $P_+>0$ that enhances the unpolarized cross sections.

Light neutralinos with a large singlet component may exist
in extended supersymmetric models, but not
necessarily as the LSP. Even then  
discrimination between supersymmetric models
by the direct production of a singlino-dominated
neutralino is a realistic goal at a future
high luminosity electron-positron linear collider.

\section*{Acknowledgment}

We thank A.~Bartl and H.~Fraas for many helpful discussions and the
careful reading of the manuscript.
This work is supported by the 'Fonds zur F\"orderung der
wissenschaftlichen For\-schung' of Austria,
FWF Project No.~P13139-PHY
and by the EU TMR Project No.\ HPRN-CT-2000-00149.
S.H. is supported by the Deutsche Forschungsgemeinschaft (DFG) under
contract No.\ HE~3241/1-1.

\end{document}